\documentstyle[12pt,psfig]{article}

\begin{document}

\voffset -2cm

\setlength{\baselineskip}{18pt} 

\noindent \hfill NBI-99-11

\vspace{8mm}

\begin{center}
{\Large \bf The Transverse Structure of the Baryon Source in Relativistic 
  Heavy Ion Collisions}

\vspace{.5cm}

Alberto Polleri$^{a,\,}$\footnote{Present address:  Institute for Theoretical 
Physics, University of Heidelberg, Philosophenweg 19, D-69120 Heidelberg, 
Germany.}, Raffaele Mattiello $^a$, Igor N. Mishustin$^{a,b}$ 
and Jakob P. Bondorf$^a$.\\[.4cm] 
{\normalsize \it $^a$The Niels Bohr Institute, Blegdamsvej 17, DK-2100 
Copenhagen \O, Denmark.}\\[.05cm]
{\normalsize \it $^b$The Kurchatov Institute, Russian Scientific Center, 
Moscow 123182, Russia.}

\vspace{.5cm}

(6 April 1999)

\end{center}

\begin{abstract}
\setlength{\baselineskip}{16pt}
A direct method to reconstruct the transverse structure of the baryon source 
formed in a relativistic heavy ion collision is presented. The procedure makes
use of experimentally measured proton and deuteron spectra and assumes that 
deuterons are formed via two-nucleon coalescence. The transverse density shape
and flow profile are reconstructed for Pb+Pb collisions at the CERN-SPS. The 
ambiguity with respect to the source temperature is demonstrated and possible
ways to resolve it are discussed.
\end{abstract}

\vspace{.3cm}

\noindent PACS number(s): 25.75 -q, 25.75 -Dw, 25.75.Ld

\vspace{.2cm}

\noindent Keywords: relativistic heavy ion collisions, coalescence model,
clusters, transverse flow.

\vspace{1cm}

One of the main goals of modern nuclear physics is to understand the behavior 
of
nuclear matter under extreme conditions. This can provide a solid basis for a
reliable description of the nuclear equation of state and therefore answer
many unsolved problems in astrophysics and early universe cosmology. In
laboratory experiments there is only one way towards this goal, which is
to study high energy collisions between heavy nuclei. 

The complicated non-equilibrium dynamics associated with such collisions are 
nowadays the object of intensive theoretical investigations. There have been 
many attempts to develop a realistic approach describing the full space-time 
evolution of the system \cite{S95,G97,BETAL98}. Furthermore, due to the large 
number of produced particles and their subsequent re-scattering, a certain 
degree of local thermodynamic equilibrium might be established at an
intermediate stage of the reaction \cite{SSH93,BMETAL96,CEST97}. One can 
therefore try to use simpler models, dealing with the properly parametrised 
phase space distribution functions for different particle species. 

However, on this road one always encounters an inherent problem.
What can be measured in an experiment are only the momentum spectra of
particles. Each measurement is only a projection of the full phase space 
information on the space-time points where particles decouple, the so-called 
freeze-out region. The fundamental ambiguity that several phase space 
distributions, after summing over the freeze-out coordinates of particles, can 
lead to the same single particle spectrum is left, in principle, unresolved.
We see that one must disentangle what can be referred to as 
an {\it inversion problem}. 

Although it is clear that independent single particle spectra are not 
sufficient to resolve the ambiguity, we fortunately have at our disposal a 
promising probe,
the deuteron (light nuclear clusters, in general). Previous studies have shown
\cite{CK86} that deuteron production cross sections can be understood in terms
of the  phenomenological coalescence model. It is founded on the fact that a 
neutron and a proton can fuse into a deuteron only if their relative distance
in position and momentum space is small, on the order of the size of the 
deuteron wave function. Furthermore, because of the small binding energy 
($\, \sim$ 2 MeV), deuterons can survive break-up only when 
scatterings are rare. These two constraints reveal that deuteron production 
can only take place at freeze-out. This is the reason why, with a suitable use
of single and composite particle spectra, one can constrain the freeze-out 
phase space distribution of protons.

Another important observation made in recent experiments 
\cite{E80294,NA4497,FOPI97} is the 
development of collective effects during the reaction. The collective behavior
is revealed by increasing inverse slopes of spectra for particles of increasing
mass. This can be understood in terms of collective flow of nuclear
matter, leading to a position-dependent local velocity.

In this letter we extend our previous work \cite{PBM98}, attempting to 
reconstruct the phase space distribution of protons directly from the observed
proton and deuteron spectra, exploiting the coalescence prescription together 
with the notion of collective flow. It is assumed that one can characterise the
system at freeze out by a position-dependent collective velocity and a 
position-independent local temperature $T_0$. We are 
mainly interested in the transverse dynamics at mid-rapidity for central 
collisions of large symmetric systems. We therefore make use of a relativistic 
description of 
collective flow based on the Bjorken picture for the longitudinal expansion, 
together with a longitudinally-independent transverse velocity 
$\vec{v}_\perp(\vec{r}_\perp)$. The contraction of the four-momentum 
$p^\mu$ with the collective four-velocity $u^\mu$ can be written as \cite{R88}
\begin{equation}
p_\mu u^\mu(x) \!=\! \gamma_\perp(r_{\!\perp})\, (m_\perp \cosh(y-\eta) - 
\vec{p}_\perp \!\cdot \,\!\vec{v}_\perp\!(r_{\!\perp}))\, ,
\end{equation}
where $m_\perp = \sqrt{m^2 + p^2_\perp}$ is the transverse particle mass
while $y$ and $\eta$ are the momentum and space-time rapidities, respectively.
Choosing a transverse profile $n_p(r_\perp)$ for the local density, we 
can write the proton phase space distribution as
\begin{equation}
f_p(x,p) = (2\pi)^3\, \mbox{\Large $\mbox{\normalsize $e$}^{-\,p_\mu 
u^\mu(x)\,/\, T_0}$}\, B_p\, n_p(r_\perp)\, .
\label{PHSP}
\end{equation}
 From now on we drop the subscripts in $\gamma_\perp$, $\vec{v}_\perp$ and
$\vec{r}_\perp$, understanding that these quantities denote the
transverse degrees of freedom. In the above expression the normalisation 
constant for the Boltzmann factor in the local frame is defined as
\begin{equation}
B_p^{-1} \!=\!\! \int\!\! d^3\vec{p}
\ \mbox{\Large $\mbox{\normalsize $e$}^{-\,m_\perp \cosh y \,/\,\!T_0}$} 
\!= 4 \pi\, m^2 T_0\, K_2(m/T_0)\, ,
\label{BOLTZ}
\end{equation}
where $K_2$ is the modified Bessel function of second order.
The local density must be normalised to the measured differential multiplicity
$dN_p/dy$. To do this we first calculate the invariant momentum spectrum.
It can be obtained by integrating the phase space distribution on the
freeze-out hypersurface, using the Cooper-Frye formula 
\begin{equation}
S_p(p_\perp) = \frac{d^3\!N_p}{dy\,d^2\vec{p}_\perp} = \frac{1}{(2\pi)^3}
\!\int\! d\sigma_\mu p^\mu \, f_p(x,p)\, .
\label{CF}
\end{equation}
The form of the integration measure follows from our simplifying choice of 
freeze-out hypersurface as a sheet of constant longitudinal proper time 
$\tau = \tau_0$. We neglect longitudinal edge effects which are negligible 
for the mid-rapidity region of the spectra. We also assume a simultaneous
freeze-out in the transverse direction and neglect surface emission
at $\tau < \tau_0$. Within this assumption the integration measure assumes 
the form
\begin{equation}
p^\mu d\sigma_\mu = \tau_0\, m_\perp\, cosh(y-\eta)\, d^2\vec{r}\, d\eta\, .
\label{MEAS}
\end{equation}
After substituting this expression in eq.$\,$(\ref{CF}), the $\eta$ integration
can be done analytically. Now the rapidity density can be obtained by
integrating out the transverse momentum dependence of $S_p(p_\perp)$. Finally, 
one arrives at the normalisation condition for the local proton density
\begin{equation}
\frac{dN_p}{dy} = 2 \pi\,\tau_0 \!\int\! dr\,r\ \gamma(r)\, n_p(r)\, .
\label{NORMR}
\end{equation}
This completes the definition of the proton phase space distribution.

The deuteron phase space distribution can be calculated
on the basis of the coalescence model. In the density matrix formalism 
\cite{SY81} one obtains
\begin{equation}
f_d(x,p) = \mbox{\small $\frac{3}{8}$}\!\!\int\!\mbox{\large $\frac{d^3\vec{y}
\,d^3\vec{q}}
{(2 \pi)^3}$}\, f_p(x_+,p_+)\, f_n(x_-,p_-) \, P_d(\vec{y},\vec{q})\, ,
\end{equation}
where the factor $\mbox{\small $\frac{3}{8}$}$ accounts for the spin-isospin 
coupling of the neutron-proton pair into a deuteron state. It is assumed that 
neutrons (not measured) evolve in the same way as protons and that $f_n\!=\! 
R_{np}\,f_p$, where $R_{np}\! =\! 1\!-\!1.5$ is the neutron to proton ratio
in the source. In the following we will take $R_{np} = 1.2$. The phase space 
coordinates of the 
coalescing pair are $x_{\pm} = x \pm y/2$ and $p_{\pm} = p/2 \pm q$ while
$P_d$ is the Wigner density for the deuteron relative motion. The coalescence
prescription is greatly simplified when considering large and hot systems.
In this case one can neglect the smearing effect of the deuteron Wigner 
density in comparison to the characteristic scales of the system in 
position and momentum space \cite{BJKG77}. Then the deuteron phase space 
distribution becomes
\begin{equation}
f_d(x,p) \simeq 
\mbox{\small $\frac{3}{8}$} R_{np} \left[ f_p(x,p/2) \right]^2\, .
\label{COAL}
\end{equation}
This expression simply means that two
nucleons, each with 4-momentum $p/2$, form a deuteron with 4-momentum $p$ at 
space-time point $x$.

Repeating the same reasoning for deuterons as done above for protons and 
inserting the corresponding expressions for $f_p$ and $f_d$ into 
eq.$\,$(\ref{COAL}), one obtains the relation
\begin{equation}
n_d(r) = \lambda_d \, n_p^2(r)
\label{COALDENS}
\end{equation}
between the local densities of deuterons and protons. The proportionality 
coefficient $\lambda_d$ has dimension $L^3$ and carries information on the 
characteristic scales in the problem. Its explicit form is
\begin{equation}
\lambda_d = \frac{3}{8} R_{np} (2\pi)^3 \frac{B_p^2}{B_d}\, .
\label{LAMBDA}
\end{equation}

We now have built up the framework to 
establish the reconstruction procedure. In a fashion which holds both for 
protons and for deuterons, using eqs.$\,$(\ref{PHSP}), (\ref{CF}) and 
(\ref{MEAS}), we can calculate their invariant momentum spectra. Due to our
simple choice of freeze-out hypersurface the integrations over the space-time
rapidity and the azimuthal angle can be easily performed, leading to the 
expression \cite{SSH93}
\begin{equation}
S(p_{\!\perp}\!)\!=\!C\!
\!\int\! dr r\, K_1(\frac{\gamma(r) m_\perp}{T_0})\,
I_0(\frac{v(r) \gamma(r) p_{\!\perp}\!}{T_0})\tau_0 n(r),
\label{AMBIG}
\end{equation}
where $C = 4\pi\,B\,m_\perp$ and $K_1$ and $I_0$ are modified Bessel functions
of first and zeroth order. Here one can explicitly see the ambiguity in the
description of the individual single particle
spectrum. The two functions $v(r)$ and $n(r)$ cannot be mapped out uniquely 
from only one function as with the transverse momentum spectrum.
This is true if protons and deuterons are treated independently, but with the
link provided by the coalescence model, this ambiguity can, at least
partially, be removed. Let us introduce for convenience a new, dimensionless, 
density function
\begin{equation}
\tilde{n}(v) = \frac{r\,dr}{v\,dv}\, \tau_0\, n(r)
\label{NTILDE}
\end{equation}
and change the integration variable in eq.$\,$(\ref{AMBIG}) from $r$ to $v$. 
In this way we obtain the new expression for the momentum spectrum
\begin{equation}
S(p_\perp) = C \!\int\! dv v\, K_1(\frac{\gamma\, m_\perp}{T_0})\,
I_0(\frac{v\, \gamma\, p_\perp}{T_0})\,\tilde{n}(v)\, ,
\label{NOAMBIG}
\end{equation}
from which the one-to-one correspondence between $\tilde{n}(v)$ and 
$S(p_\perp)$ is evident, since all the functions in the integrand are 
single-valued. The normalisation condition (\ref{NORMR}) for $n_p(r)$, due to the
definition (\ref{NTILDE}), now becomes
\begin{equation}
\frac{dN_p}{dy} = 2 \pi \!\int\! dv\,v\ \gamma\ \tilde{n}_p(v)\, ,
\end{equation}
together with the analogous one
for $\tilde{n}_d$. It is instructive to consider the limit $T_0 \rightarrow
0$ in eq.$\,$(\ref{NOAMBIG}). Using
the asymptotic expression for large argument for the Bessel functions $K_1$
and $I_0$ and performing the integration with the saddle point method, one 
obtains that $v = p_\perp/m_\perp$ and $m_\perp = m \gamma$. The transverse 
momentum spectrum is simply expressed through $\tilde{n}$ as
\begin{equation}
S(p_\perp) \sim \frac{m}{m_\perp^3}\, \tilde{n}(p_\perp/m_\perp)\, .
\label{TLIMIT}
\end{equation}
One can easily find $\tilde{n}$ when the observed spectrum has an
exponential shape with inverse slope $T_*$ as
\begin{equation}
S^{exp}(p_\perp) \sim \frac{m_\perp}{(m\,T_*)^{3/2}} \mbox{\Large $\mbox{\normalsize
 $e$}^{ - (m_{\!\perp} - \,m)/T_*}$}\, .
\end{equation}
Expressing the momentum variable in terms of velocity, $p_\perp = m v \gamma$,
we obtain
\begin{equation}
\tilde{n}(v) \sim 
\gamma^4\,b_0^{3/2} \mbox{\Large $\mbox{\normalsize $e$}^{ -
\,b_0\,(\gamma -1)}$}\, ,
\label{SHAPELIM}
\end{equation}
with $b_0 = m/T_*$. Observe that the exponent behaves as 
$\exp\,(-\,b_0\,v^2/2)$ for small $v$ and as $\exp\,(-\,b_0\,\gamma)$ for 
large $v$. In the limiting case $T_0 \rightarrow 0$ the function
$\tilde{n}$ is therefore uniquely determined from the
spectrum. A finite temperature introduces an inevitable intrinsic 
smearing, which precludes an exact determination of $\tilde{n}$ for a single
particle species.

The role of the coalescence model is now to establish the link 
between proton and deuteron spectra. Substituting in eq.$\,$(\ref{COALDENS})
the definition of $\tilde{n}$ from eq.$\,$(\ref{NTILDE}), both for protons and
for deuterons, we obtain the simple differential equation
\begin{equation}
\tilde{n}_d(v) = \frac{\lambda_d}{\tau_0}\, \frac{v\, dv}{r\, dr}
\ \tilde{n}_p^2(v)\, ,
\end{equation}
which can be directly integrated and leads to the closed solution
\begin{equation}
r^2 = 2\, \frac{\lambda_d}{\tau_0} \int_0^{v}\!\! du\,u\, 
\frac{\tilde{n}_p^2(u)}{\tilde{n}_d(u)}\,.
\label{FLOWREC}
\end{equation}
Therefore, by independently extracting the functions $\tilde{n}_p$ and 
$\tilde{n}_d$ from the observed momentum spectra $S_p(p_\perp)$ and 
$S_d(p_\perp)$, we can find the function $r(v)$ by a simple numerical 
integration. Inverting the obtained function as $r(v) \rightarrow v(r)$,
we obtain the collective velocity profile. Substituting it in 
eq.$\,$(\ref{NTILDE}), we finally obtain the local proton density
\begin{equation}
n_p(r) = \frac{v(r)}{r}\ \frac{dv(r)}{dr}\ \frac{\tilde{n}_p(v(r))}{\tau_0}\, .
\label{DENSREC}
\end{equation}

We now explore more closely the content of eqs.$\,$(\ref{FLOWREC}) and
(\ref{DENSREC}). 
Taking the limit for small transverse velocity in eq.$\,$(\ref{FLOWREC}),
one finds that $v(r) \simeq H\, r$
and the dimensionful scale in position space is set by the ``Hubble constant''
\begin{equation}
H = \left(\frac{\tau_0\, \tilde{n}_d(0)}
{\lambda_d\, \tilde{n}_p^2(0)}\right)^{\!\!1/2} \!\!\! =
\left(\frac{\tau_0}{3\, R_{np}} 
\left(\frac{m\, T_0}{\pi}\right)^{\!\!\!3/2}
\!\!\frac{\tilde{n}_d(0)}
{\tilde{n}_p^2(0)}\right)^{\!\!1/2} \!\!\!.
\label{SCALE}
\end{equation}
The r.h.s. of this expression is obtained from eqs.$\,$(\ref{BOLTZ}) and
(\ref{LAMBDA}), using the asymptotic expression for large argument for the 
Bessel function $K_2$ in the Boltzmann factors. Using this result and taking 
the limit for $r \rightarrow 0$ in eq.$\,$(\ref{DENSREC}), we obtain the 
proton spatial density at the origin
\begin{equation}
n_p(0) \simeq \frac{1}{3\, R_{np}} \left(\frac{m\, T_0}{\pi}\right)^{\!\!\!3/2}
\!\frac{\tilde{n}_d(0)}{\tilde{n}_p(0)}\, .
\label{CONSTRAINT}
\end{equation}
This result will be used in the following discussion.

\begin{figure}
\centerline{\psfig{figure=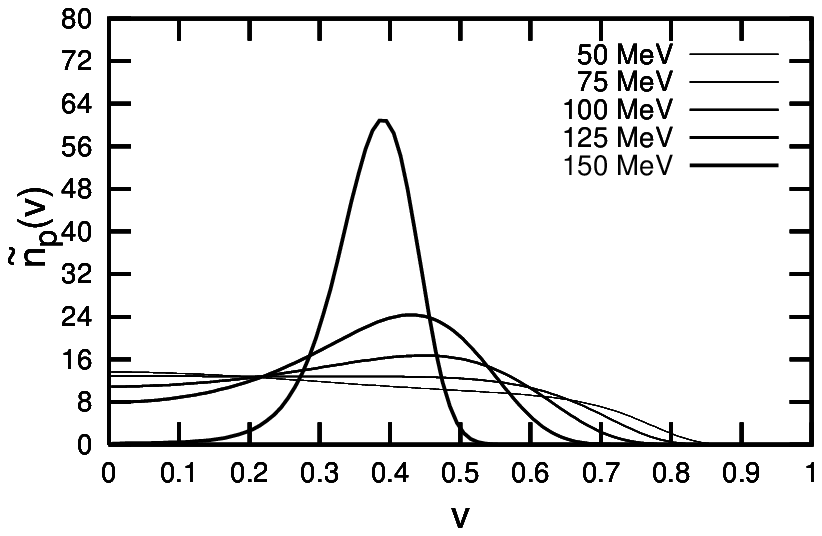,width=15cm,angle=-90}} 
\protect\caption{Profile $\tilde{n}_p$, characterized by the 
parameters shown in tab.$\,$\ref{TABLE}.}
\label{PPROF}
\end{figure}

\begin{figure}
\centerline{\psfig{figure=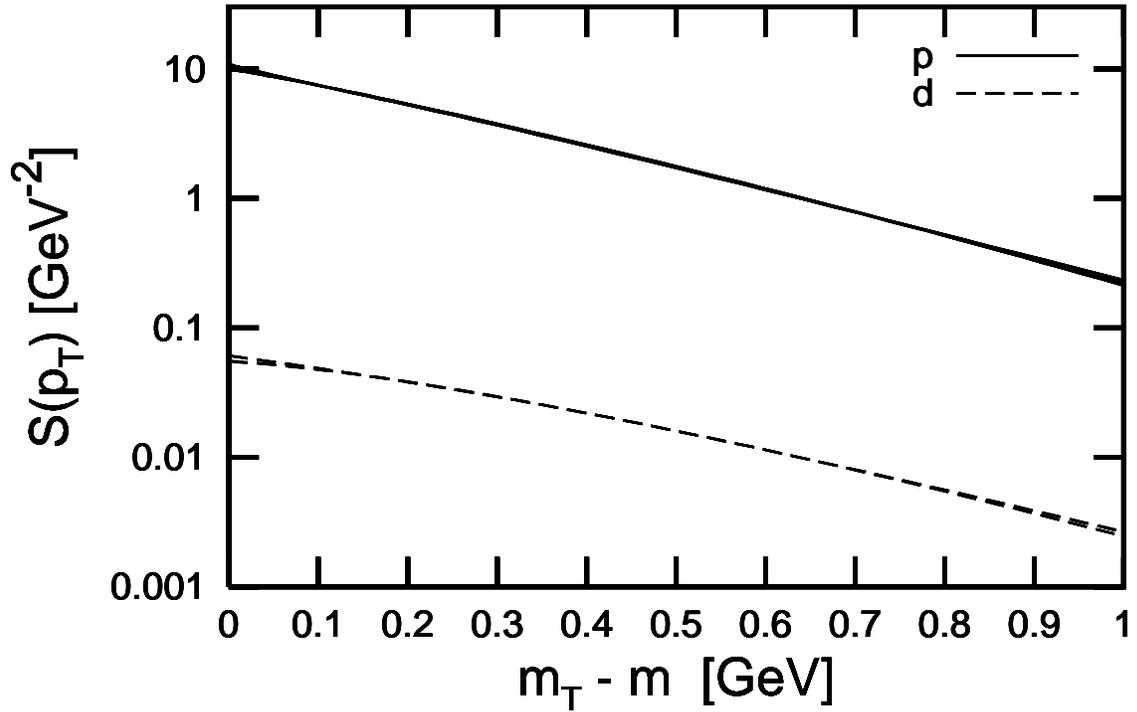,width=15cm,angle=-90}}
\protect\caption{Transverse momentum spectra for protons (top) and deuterons 
     (bottom), plotted as functions of $m_\perp - m$. They are obtained by
     fitting the NA44 data \protect\cite{NA4499}, using the trial functions
     of eq.$\,$(\ref{NOAMBIG}) with the parameters shown in 
     Tab.$\,$\ref{TABLE}. Calculated curves corresponding to different
     temperatures are superimposed and indistinguishable.}
\label{SPEC}
\end{figure}

The described procedure to reconstruct the proton phase space distribution can
now be applied to the transverse momentum spectra, measured by the NA44 
collaboration at the CERN-SPS, for Pb+Pb collisions at $158$ A$\,$GeV 
\cite{NA4499}. Using these data, we fitted $\tilde{n}_p$ and $\tilde{n}_d$, 
using eq.$\,$(\ref{NOAMBIG}) for protons and deuterons and assuming different 
values for $T_0$. Guided by 
eq.$\,$(\ref{SHAPELIM}) we chose a rather flexible form of the 
profile function, characterized by three parameters, $k$, $a$ and $b$, as
\begin{equation}
\tilde{n}(v) \sim \gamma^k \, \mbox{\Large $\mbox{\normalsize 
$e$}^{ - (a-b/2)\,v^2 - b\,(\gamma - 1) }$}\, .
\label{PARAMETRIZE}
\end{equation}
The exponential factor has now the limiting behavior $\exp\,(- a\,v^2)$ for
small $v$, while $\exp\,(- b\,\gamma)$ for large $v$. We separated
the scales $a$ and $b$ for small and large $v$, in order to have more
freedom in fitting the curvature of the spectrum.
The parameters $k_p$, $a_p$, $b_p$, $k_d$, $a_d$, 
$b_d$, are extracted from the experimental data with a Monte Carlo search
minimising
\begin{equation}
\chi^2 = \sum_j \left( \frac{S^{exp}(p_j) - S^{theo}(p_j)}{S^{exp}(p_j)} 
\right)^2\, .
\end{equation}
The only constraint imposed was $\delta_d > 2\delta_p$: because of our choice 
of $\tilde{n}$ in eq.$\,$(\ref{PARAMETRIZE}), the 
integrand in eq.$\,$(\ref{FLOWREC}) is exponentially divergent for $v 
\rightarrow 1$, thereby giving the limit $v(r) \rightarrow 1$ for $r 
\rightarrow \infty$.
The values obtained in this way are listed in Tab.$\,$\ref{TABLE} and the 
reconstructed function $\tilde{n}_p$ is shown in Fig.$\,$\ref{PPROF}. We do not
show the deuteron profile since it is very similar to the proton one. The 
fitted spectra, normalised as $dN_p/dy = 22$ and $dN_d/dy = 0.3$, are shown in 
Fig.$\,$\ref{SPEC}. One clearly sees that the calculated spectra for different
temperatures are indistinguishable from one another. On the other hand, the
reconstructed profiles $\tilde{n}$ are very different for
different temperatures. They are very much peaked at large temperatures and 
become broad as the temperature drops. This is simple to understand, since
a lower temperature introduces less momentum spread than a higher one and 
$\tilde{n}$ compensates for that by broadening. One can notice that the 
high-temperature profiles have a structure resembling a blast wave \cite{SR79}
where most particles have approximately the same velocity.

\begin{figure}
\centerline{\psfig{figure=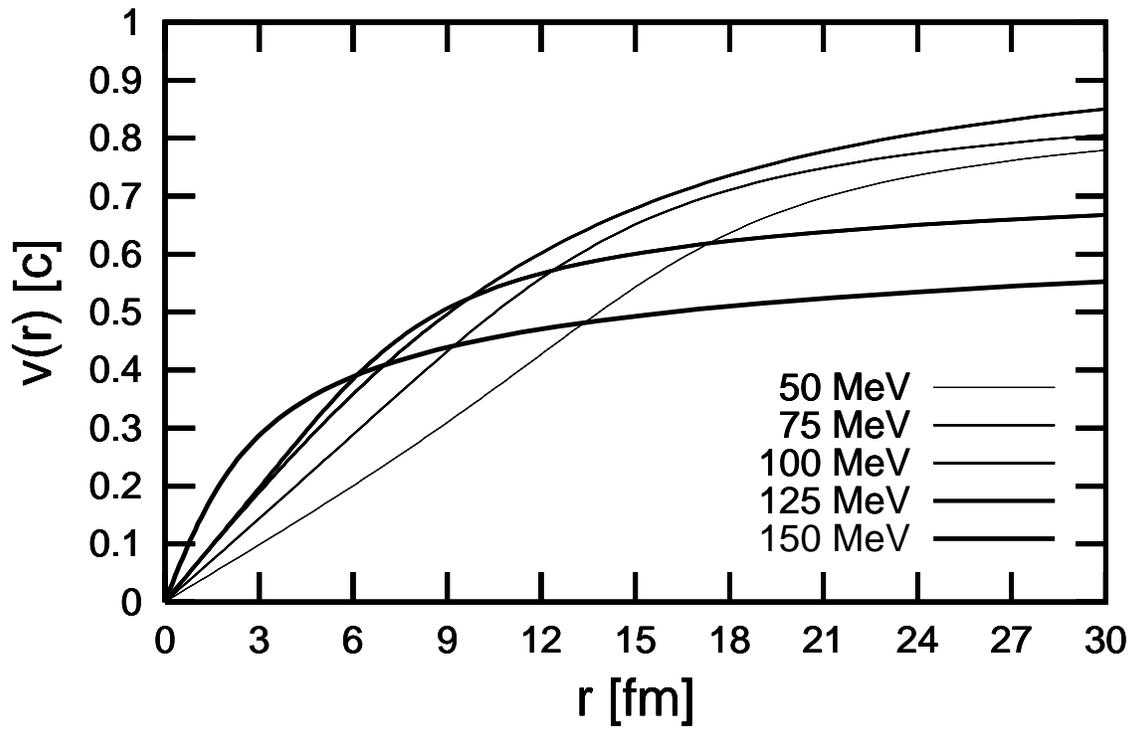,width=15cm,angle=-90}}
\protect\caption{Reconstructed transverse flow velocity fields for different
     temperatures. The profiles shows a strong rise at small $r$,
     levelling off at large $r$.}
\label{FLOW}
\end{figure}

In all calculations the freeze-out time was fixed at $\tau_0 = 10$ fm/c 
\cite{BBGH97}. After numerical integration of eq.$\,$(\ref{FLOWREC}), we 
obtained the function $v(r)$ shown in Fig.$\,$\ref{FLOW} for different 
temperatures. It shows a linear rise at small $r$ and saturates
for large $r$. The velocity profiles clearly depend on the temperature chosen.

\begin{figure}
\centerline{\psfig{figure=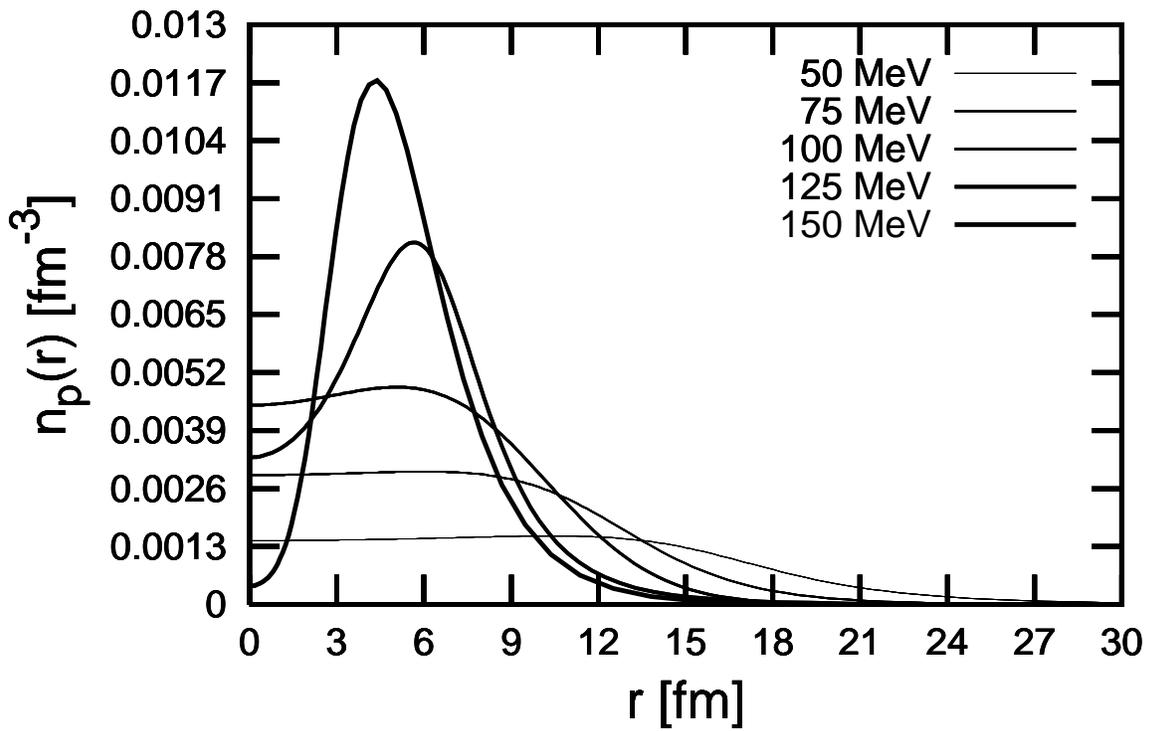,width=15cm,angle=-90}}
\protect\caption{Reconstructed local density of protons as a function of the 
         transverse coordinate. The resulting profile shows a shell-like
         structure at high temperature, turning into a more uniform behavior
         at smaller $T_0$.}
\label{DENS}
\end{figure}

Making use of eq.$\,$(\ref{DENSREC}) we also obtained the local proton density,
plotted in Fig.$\,$\ref{DENS}. One can observe a behavior similar to 
$\tilde{n}_p$. At high temperature the density shows a shell-like structure 
which disappears as the temperature is lower. It should be emphasised that the
shapes of $v(r)$ and $n_p(r)$ are quite sensitive to the curvature of the
momentum spectra.

 From the plots of $v(r)$ in Fig.$\,$\ref{FLOW}, one might have the impression
that flow is
somehow stronger at high temperature. This is not the case, since the apparent
flattening of $v(r)$ at small $r$ is accompanied by a higher saturation value
and a corresponding larger extension of $n_p(r)$. To see this more 
quantitatively we calculated the mean-square value for the velocity field.
The mean, taken with respect to the proton density in the global frame defined
by $\rho_p(r) = \gamma(r)\,n_p(r)$, is calculated as
\begin{equation}
\langle\,v^2\,\rangle = \left(\frac{dN_p}{dy}\right)^{-1}\!\!\int\! 
d^2r\ v^2(r)\ \tau_0\,\rho_p(r)\, ,
\end{equation}
and the results are listed in Tab.$\,$\ref{TABLE} for different
temperatures. As one can see, $\langle\,v^2\,\rangle^{1/2}$ indeed grows
with decreasing $T_0$.

The broadening of $n_p$ as the temperature decreases is dictated by the
conservation of the total phase space volume. The particular value chosen for 
$\tau_0$ also affects the resulting transverse extension of the source. All of
this is evident after examining eq.$\,$(\ref{SCALE}). 
A large temperature results in a large $H$, {\it i.e.} in a smaller transverse 
extension scale $H^{-1}$. As a consequence the average proton density is
larger. This is explicit in eq.$\,$(\ref{CONSTRAINT}). To quantify the 
transverse size of the source we have also calculated the mean-square 
transverse radius
\begin{equation}
\langle\,r^2\,\rangle = \left(\frac{dN_p}{dy}\right)^{-1}\!\!\int\! 
d^2r\ r^2\ \tau_0\,\rho_p(r)\, .
\end{equation}
The resulting values of $\langle\,r^2\,\rangle^{1/2}$ are listed 
in Tab.$\,$\ref{TABLE}. One can observe the increasing mean transverse size
with decreasing temperature. It is therefore necessary to know $T_0$ precisely
in order to determine the system size. This information cannot be  extracted 
solely from proton and deuteron spectra. Contrary to what is commonly done, 
source radii {\it cannot} be extracted from the $d/p^2$ ratio (usually assumed
to be inversely proportional to the source volume) unless $T_0$ is known or
further assumptions are made. One possibility relies on a rough estimate of 
the characteristic value of $n_p$ at freeze-out. Taking the total hadron 
number density to be $\sim \rho_0/3$, where $\rho_0 = 0.17\ \mbox{fm}^{-3}$ is
the nuclear saturation density, with an average hadron-hadron cross section of
$\sim 30$ mb, we have a mean free path of $\sim 6$ fm. This is long enough to 
reach the 
freeze-out conditions, especially for a rapidly expanding system. Since the 
ratio of all hadrons ($p$, $n$, $\pi$, $K$, $\dots\,$) to protons is $\sim 15$
for Pb+Pb collisions at $158$ A$\,$GeV \cite{NA4998}, we can estimate that
\begin{equation}
n_p \simeq \frac{1}{15}\times\frac{1}{3}\,\times\,0.17\ \mbox{fm}^{-3}
\simeq 0.0038\ \mbox{fm}^{-3}\, .
\end{equation}
With this number, one can say that our reconstruction procedure
suggests the preferred values of $T_0 \simeq 100$ MeV and 
$\langle\,r^2\,\rangle^{1/2} \simeq 9.2$ fm, as follows from Fig.$\,$\ref{DENS}
and Tab.$\,$\ref{TABLE}. This is also consistent with the temperature 
dependence in eq.$\,$(\ref{CONSTRAINT}). Comparison with microscopic
simulations of the freeze-out distributions at AGS energies 
\cite{BMABC95,MSSG96} and at SPS energies \cite{S96} also shows 
that the temperature around 100 MeV would be preferable. To make a definite 
statement, one must finally keep in mind that the average source size is 
influenced by the choice of $\tau_0$. A larger (smaller) value would result in
a smaller (larger) average size. Internal consistency requires that $\tau_0$
is large enough so that the flow field has time to develop and the source to
expand transversely during the reaction. Details about these isuues can
only be discussed within a dynamical calculation, unless an independent
measurement of $\tau_0$ is available.

It is clear that to resolve the remaining ambiguity in the source temperature
one needs some additional experimental information. Recently, $\pi\,\pi$ HBT 
correlations data were used as a constraint \cite{SH99}. Since pions may 
freeze-out in a different way than protons, it would be even better to
consider $p\,p$ HBT 
correlations, although they are much more sensitive to final state 
interactions than pions. We would also like to mention that the neutron
phase space distribution, here assumed to be proportional to the proton one,
could be also reconstructed with a similar procedure for triton 
spectra. Furthermore, heavier clusters are more sensitive to collective flow
than to temperature \cite{MSSG96}, providing additional constraints.

In conclusion, we have demonstrated how the proton phase space distribution at
freeze-out can be reconstructed from the measured transverse momentum spectra 
of protons and deuterons. The calculations, made with several simplifying 
assumptions, show that the proposed method gives meaningful results, although
the ambiguity with respect to the temperature of the source remains. Two
modifications will make this approach more realistic. First, the deuteron size
should be explicitly included in the calculation. The wave function's tail 
plays a quantitatively important role, especially at high energies, when the 
large number 
of produced particles forces nucleons to be far apart at freeze-out. Second,
the freeze-out hypersurface itself is more complicated than a constant proper 
time hyperbola. A correct description of early evaporation of fast 
particles is needed in order to have the high momentum part of the spectra 
under control. Nevertheless the described procedure provides a basis for 
more quantitative analyses, possibly including  HBT correlations for protons.

\vspace{3mm}

The authors would like to thank Andrew Jackson for many valuable and 
constructive suggestions and Ian Bearden, Jens J{\o}rgen Gaard{\o}je, 
Allan Hansen and the NA44 group for a very fruitful collaboration
and for providing their preliminary data for our analysis. This 
work was supported in part by I.N.F.N. (Italy).

\newpage

\begin{table}
\protect\caption{Fitted parameters characterizing the profiles $\tilde{n}_p$ 
and $\tilde{n}_d$ for different values of the source temperature $T_0$. 
Root-mean-square values for velocity and transverse radius are also listed.}
\begin{center}
\begin{tabular}{|c|c|c|c|c|c|}\hline
$T_0$ & $k_p$ & $a_p$ & $b_p$ & $\langle\,v^2\,\rangle^{1/2}$ 
& $\langle\,r^2\,\rangle^{1/2}$\\ \cline{2-4}
(MeV) & $k_d$ & $a_d$ & $b_d$ & & (fm) \\ \hline\hline
 50 &  91.56 &  47.63 &   52.40 &  0.54 & 17.60 \\ \cline{2-4}
    & 165.29 &  83.55 &  107.77 &       & \\ \hline
 75 & 108.50 &  54.37 &   68.91 &  0.51 & 11.93 \\ \cline{2-4}
    & 214.21 & 106.65 &  146.04 &       & \\ \hline
100 &  75.86 &  34.19 &   66.45 &  0.48 &  9.20 \\ \cline{2-4}
    & 146.62 &  67.85 &  133.62 &       & \\ \hline
125 &  68.48 &  23.18 &  105.00 &  0.44 &  8.49 \\ \cline{2-4}
    &  67.24 &   9.78 &  211.05 &       & \\ \hline
150 & 172.89 &  17.47 &  602.83 &  0.39 &  7.02 \\ \cline{2-4}
    & 307.96 &  31.06 & 1211.85 &       & \\ \hline
\end{tabular}
\end{center}
\label{TABLE}
\end{table}


\begin{thebibliography}{25}

\bibitem{S95}
H. Sorge,
Phys. Rev. {\bf C52}, 3291 (1995)

\bibitem{G97}
K. Geiger,
Comput. Phys. Commun. {\bf 104}, 70 (1997)

\bibitem{BETAL98}
S.A. Bass {\it et al.},
Prog. Part. Nucl. Phys. {\bf 41}, 225 (1998).

\bibitem{SSH93}
E. Schnedermann, J. Sollfrank and U. Heinz,
Phys. Rev. C {\bf 48}, 2462 (1993).

\bibitem{BMETAL96}
P. Braun-Munzinger, J. Stachel, J.P. Wessels and N. Xu,
Phys. Lett. {\bf B344}, 43 (1995); {\bf B365}, 1 (1996)

\bibitem{CEST97}
J. Cleymans, D. Elliott, H. Satz and R.L. Thews,
Z. Phys. {\bf C74}, 319 (1997)

\bibitem{CK86}
L.P. Csernai and J.I. Kapusta,
Phys. Rep. {\bf 131}, 223 (1986).

\bibitem{E80294}
T. Abbott {\it et al.}, E802 Collaboration,
Phys. Rev. C {\bf 50}, 1024 (1994).

\bibitem{NA4497}
I. Bearden {\it et al.}, NA44 Collaboration, 
Phys. Rev. Lett. {\bf 78}, 2080 (1997).

\bibitem{FOPI97}
W. Reisdorf {\it et al.}, FOPI Collaboration,
Nucl. Phys. A {\bf 612}, 493 (1997).

\bibitem{PBM98}
A. Polleri, J.P. Bondorf and I.N. Mishustin,
Phys. Lett. B {\bf 419}, 19 (1998).

\bibitem{R88}
P.V. Ruuskanen,
Acta Phys. Pol. B {\bf 18}, 551 (1987).

\bibitem{SY81}
H. Sato and K. Yazaki,
Phys. Lett. B {\bf 98}, 153 (1981).

\bibitem{BJKG77}
R. Bond, P.J. Johansen, S.E. Koonin and S. Garpman,
Phys. Lett. B {\bf 71}, 43 (1977).

\bibitem{NA4499}
I. Bearden {\it et al.}, NA44 Collaboration, 
{\it private communication}.

\bibitem{SR79}
P.J. Siemens and J.O Rasmussen,
Phys. Rev. Lett. {\bf 42} 880 (1979).

\bibitem{BBGH97}
H.W. Barz, J.P. Bondorf, J.J. Gaardhoje and H. Heiselberg,
Phys. Rev. {\bf C56}, 1553 (1997)

\bibitem{NA4998}
H. Appelsh\"auser {\it et al.},
Nucl. Phys. A {\bf 638}, 91c (1998).

\bibitem{BMABC95}
L.V. Bravina, I.N. Mishustin, N.S. Amelin, J.P. Bondorf and L.P. Csernai,
Phys. Lett. {\bf B354}, 196 (1995); Heavy Ion Phys. {\bf 5} 455 (1997).

\bibitem{MSSG96}
R. Mattiello {\it et al.},
Phys. Rev. Lett. {\bf 74}, 2180 (1995);
R. Mattiello, H. Sorge, H. St\"{o}cker and W. Greiner,
Phys. Rev. C {\bf 55}, 1443 (1997).

\bibitem{S96}
H. Sorge,
Phys. Lett. {\bf B373}, 116 (1996).

\bibitem{SH99}
R. Scheibl and U. Heinz,
Phys. Rev. {\bf C59}, 1585 (1998).

\end{thebibliography}
\end{document}